# Magnetic anisotropy of $La_{0.7}Sr_{0.3}MnO_3$ nanopowders

I. Radelytskyi (1), P. Dłużewski (1), V. Dyakonov (1 and 2), P. Aleshkevych (1), W. Kowalski (1), P. Jarocki (1), H. Szymczak (1)

((1) Institute of Physics, PAS, Warsaw, Poland, (2) A.A.Galkin Donetsk Physico-Technical Institute, NASU, Donetsk, Ukraine)

The magnetic anisotropy of $La_{0.7}Sr_{0.3}MnO_3$ nanopowders was measured as a function of temperature by the modified singular point detection technique. In this method singularities indicating the anisotropy field were determined analyzing *ac* susceptibility data. The observed relationship between temperature dependence of anisotropy constant and temperature dependence of magnetization was used to deduce the origin of magnetic anisotropy in the nanopowders. It was shown that magnetic anisotropy of $La_{0.7}Sr_{0.3}MnO_3$ nanopowder is determined by two-ion (dipolar or pseudodipolar) and single-ion mechanisms.

## 1.Introduction

Manganites of perovskite structure have attracted much attention due to discovery of the colossal magnetoresistance effect ([1], see also review papers [2-4]). Among them especially interesting is $La_{0.7}Sr_{0.3}MnO_3$ (LSMO) due to its relatively high the Curie temperature (about 370 K) and high value of magnetization at room temperature. These parameters in combination with a high spin polarization of charge carriers make them interesting for practical applications. Recently, we have performed detailed studies of magnetic, resonance and transport properties of $La_{0.7}Sr_{0.3}MnO_3$ nanopowders [5]. It has been shown that both the Curie temperature and magnetization decrease with reducing the particle size. The decrease of magnetization and others intrinsic parameters of the nanoparticles have its origin in surface effects. The broken symmetry at the surface is also a source of the surface anisotropy which determines the effective anisotropy of the nanoparticles. The effective magnetic anisotropy essentially controls the hysteretic behavior of ferromagnets and, consequently, determines most of their parameters (e.g. coercivity, permeability, energy of magnetic domain walls) important for practical applications. Therefore, the understanding of magnetic anisotropies in nanoparticles is of crucial importance for the development of various magnetic devices. Although the magnetic anisotropy is one of the fundamental intrinsic properties of magnetic materials according to our best knowledge the magnetic anisotropy characteristics of manganite nanopowders have not been systematically investigated.

In the present paper, we report on the magnetic anisotropy of $La_{0.7}Sr_{0.3}MnO_3$ nanopowders which are investigated by *ac* SQUID technique. The paper is a continuation of the studies described in [5]. Since structural and magnetic properties of $La_{0.7}Sr_{0.3}MnO_3$ nanopowders were discussed in [5] we limit ourselves only to investigation of their magnetic anisotropy.

## 2. Experimental details (Materials and methods)

The $La_{0.7}Sr_{0.3}MnO_3$ nanopowders were prepared by a simple chemical co-precipitation procedure described earlier [5]. Using this method the nanopowders with the average particle sizes of 17 nm, 27 nm (determined by X-ray diffraction measurements) and 94 nm (determined by electron microscope observations) were obtained for 600, 700 and 900 $^o$C annealing temperatures, respectively. The $La_{0.7}Sr_{0.3}MnO_3$ nanopowders were pressed under the pressure of 0.2 GPa into pellets. Room temperature x-ray diffraction measurements have showed the formation of perovskite-like homogeneous single phase with a rhombohedral distortion (space group R-3c). The magnetization measurements performed as a function of temperature have indicated for all the samples ferromagnetic- like magnetic ordering with Curie temperature ($T_C$) increasing with increase of nanoparticle size. Structural characterization of the samples was performed by X-ray diffraction (XRD) and high-resolution transmission electron microscopy (HR-TEM).

The magnetic anisotropy of the $La_{0.7}Sr_{0.3}MnO_3$ nanopowders was investigated by the use of the singular point detection technique developed by Asti and Rinaldi [6]. This powerful technique was developed for the study of singularities in magnetization process when it reaches saturation at the anisotropy field $H_{an}$. This singular point is determined by observing the successive derivatives $d^nM/dH^n$. The measurements are usually performed in pulsed magnetic field. In this paper a modification of the singular point detection technique proposed by Turilli [7] was used. The modification consists in detection of the singular points analyzing the *ac* susceptibility in longitudinal geometry (*ac* field parallel to the static field *H*). This method was successfully applied to study magnetocrystalline anisotropy of polycrystalline $Gd_5Ge_2Si_2$ [8]. The *ac* susceptibility measurements were carried out by a 7T-superconducting quantum interference device SQUID magnetometer (Quantum Design MPMS – XL).

## 3. Structure characterization

For detailed structural characterization the samples annealed at $600^0C$ and $900^0C$ were chosen. The samples were investigated with a use of Titan Cubed image Cs corrected transmission electron microscope operating at 300 kV acceleration voltage. Specimens for TEM observations were prepared in two step process. First the nanoparticles were mixed with methanol and a droplet of the resulting suspension was deposition onto carbon holey film and drying. To prevent the agglomeration of nanoparticles the ultrasonic treatment was applied. From an analysis of microscope images the size of the nanoparticles was determined. It was found that the mean size of nanoparticles was 20 nm and 94 nm for the samples annealed at $600^0C$ and $900^0C$, respectively. The TEM observations did not reveal core-shell particles. It is in contrast to the indirect measurements based on electron magnetic resonance technique [9],[10] and to our previous conclusion [5] based only on magnetic measurements. High-resolution images revealed that the nanoparticles had mono-crystalline structure.

The analysis of lattice fringes patterns (see Figs 1 and 2), for some of the particles, confirmed perovskite-like crystal structure with a rhombohedral distortion (space group R-3c) [5].

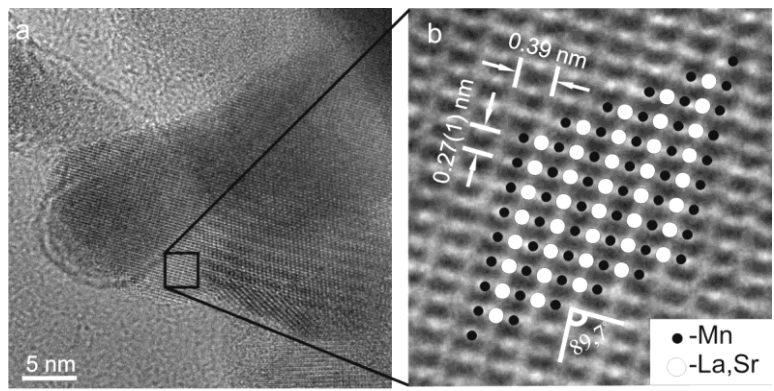

**Fig. 1.** a) The HR-TEM image of the $La_{0.7}Sr_{0.3}MnO_3$ sample annealed at 600 °C and b) the enlarged rectangular region of a) showing projection of the perovskite-like crystal structure with a rhombohedral distortion in the <0-11> direction (the oxygen atoms are not shown)

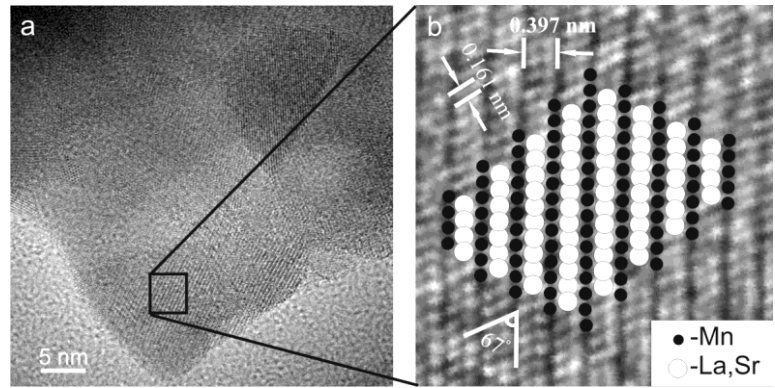

**Fig. 2.** a) The HR-TEM image of the $La_{0.7}Sr_{0.3}MnO_3$ sample annealed at 900 °C and b) the enlarged rectangular region of a) showing projection of the perovskite-like crystal structure with a rhombohedral distortion structure in the <1-13> direction (the oxygen atoms are not shown)

## 4. Magnetic measurement

Fig. 3 shows the magnetic field dependences of magnetization at different temperatures for $La_{0.7}Sr_{0.3}MnO_3$ nanopowder annealed at the temperature 900 °C. The values of magnetization measured at low temperatures weakly depend on temperature (see inset in Fig.3). The dependences of the magnetization on the magnetic field for the samples $La_{0.7}Sr_{0.3}MnO_3$ annealed at 600 and 700 °C for further analysis were taken from [5].

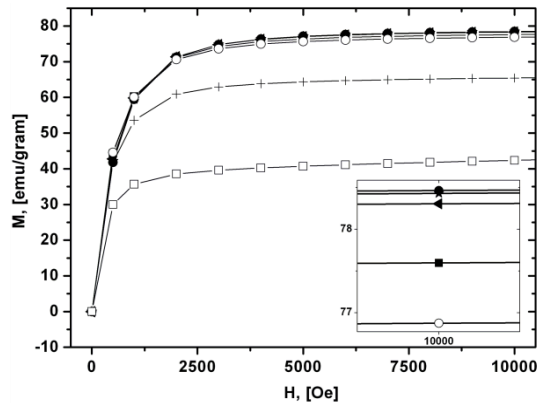

**Fig. 3.** The magnetic field dependence of magnetization for the $La_{0.7}Sr_{0.3}MnO_3$ sample annealed at 900 °C, at experimental temperatures: 5 (circle), 10 (star), 20 (triangle), 50 (square), 70 (open circle), 200 (cross) and 300 K (open square). The inset shows the $M(H)$ curves near $H = 10\,000$ Oe for several temperatures: $T = 5, 20, 50$ and $70$ K

The saturation magnetization was determined from the extrapolated magnetization data for $La_{0.7}Sr_{0.3}MnO_3$ nanopowders using the following relationship [11]:

$$M_H = M_s(T)[1 - x/H - y/H^2], \qquad (1)$$

where $M_H$ is the measured magnetization in the field $H$, $M_s(T)$ is the saturation magnetization at temperature $T$ and $x, y$ are the constants.

The extrapolated values of magnetization $M_s(T)$ for the $La_{0.7}Sr_{0.3}MnO_3$ nanopowder annealed at 900 °C are displayed in Fig. 4 complementing such the results presented in [5].

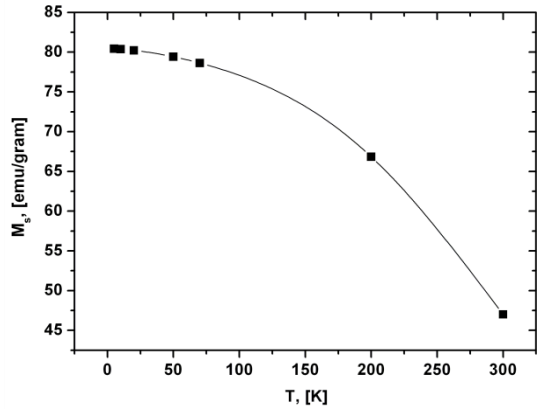

**Fig. 4.** The temperature dependence of the saturation magnetization for the $La_{0.7}Sr_{0.3}MnO_3$ nanopowder annealed at 900 °C

Fig. 5 presents the magnetic *ac* susceptibility, $\chi'$, as a function of temperature for the $La_{0.7}Sr_{0.3}MnO_3$ nanopowders. The presented results clearly demonstrate a reduction of the magnetic susceptibility with decreasing the nanoparticles size. Such behavior of $\chi'$ is expected to appear due to the presence of surface effects. The same effects are also seen in Fig. 6, where the magnetic susceptibility dependences on magnetic field measured at different temperatures are plotted for nanopowders with different grain sizes. It is seen that the applied magnetic field decreases the size effect. The observed effect suggests existence of noncollinear magnetic structure at the surface of the particles. External magnetic field transforms continuously this structure to collinear one. Such transformation may also be induced by temperature. Probably, the temperature- induced transformations of surface structure are responsible for peculiarities seen in Fig. 5.

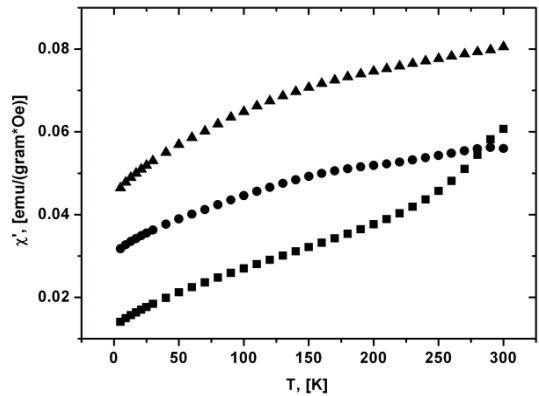

**Fig. 5.** The temperature dependence of the magnetic *ac* susceptibility in the constant field $H = 0$ and oscillation field $h(ac) = 1$ Oe at frequency of 10 Hz for the $La_{0.7}Sr_{0.3}MnO_3$ nanopowders annealed at 600 °C (square), 700 °C (circle) and 900 °C (triangle)

Electron microscopy measurements indicate that studied nanoparticles are structurally homogeneous without any core-shell structure. At the same time performed magnetic measurement unambiguously show nonhomogeneous magnetic structure with noncollinear magnetic structure at magnetic shell. External magnetic field transforms this noncollinear magnetic structure into collinear one and in this way suppress magnetic core-shell structure.

The modified singular point detection technique was applied to determine the magnetic anisotropy field of nanopowders. Figure 7 shows several examples of the $d\chi'/dH$ dependence versus the internal field $H$ ($H = H_{ext} - NM_s$, where $N$ is the demagnetizing factor of the sample) measured at various temperatures (not all of the obtained results are presented in Fig. 7) and for various grain sizes. Singularities observed in these Figures determine the anisotropy field $H_{an}$. The anisotropy field is related to the second order anisotropy constant $K_2$ through the relation:

$$H_{an} = 2K_2/M \qquad (2)$$

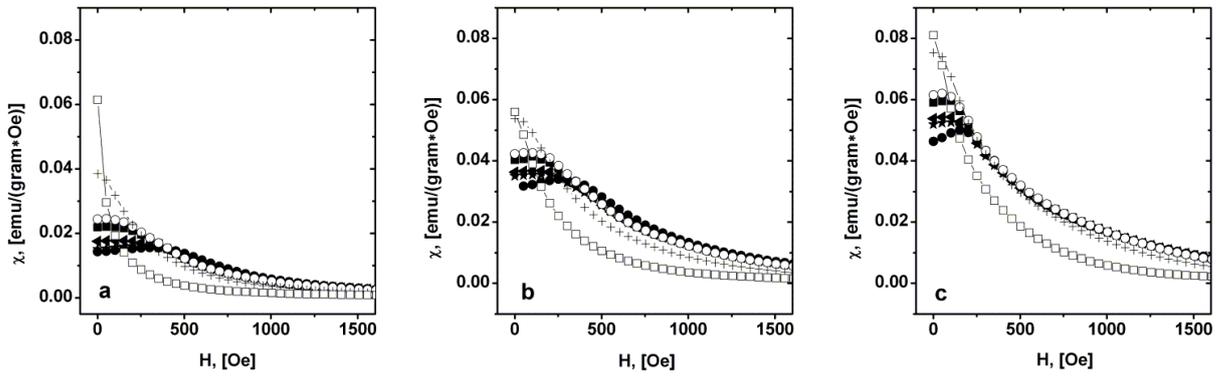

**Fig. 6 (a,b,c).** The magnetic field dependences of *ac* susceptibility (at frequency of 10 Hz) for the $La_{0.7}Sr_{0.3}MnO_3$ samples, annealed at 600 (a), 700 (b) and 900 °C (c), at the experimental temperatures : 5 (circle), 10 (star), 20 (triangle), 50 (square), 70 (open circle), 200 (cross) and 300 K (open square)

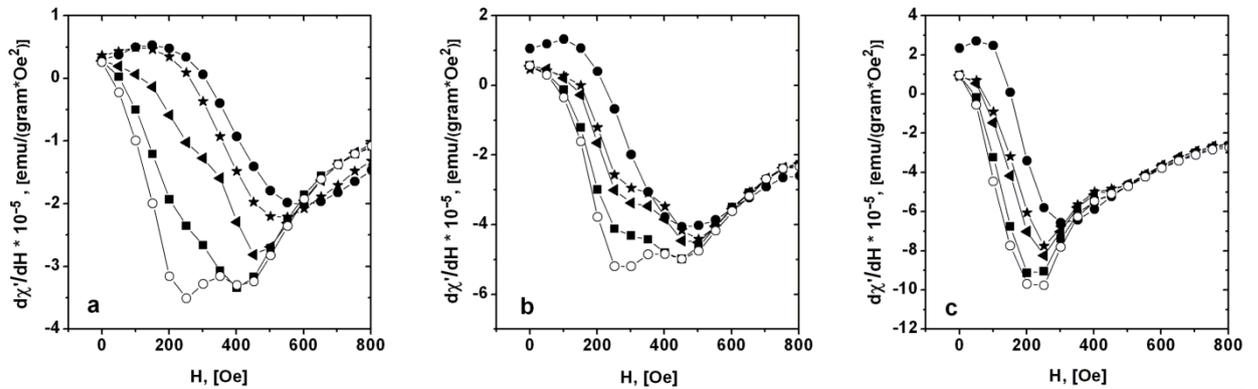

**Fig. 7 (a,b,c).** The magnetic field dependences of the derivative of magnetic *ac* susceptibility for the $La_{0.7}Sr_{0.3}MnO_3$ samples, annealed at 600 (a), 700 (b) and 900 °C (c), at the experimental temperatures : 5 (circle), 10 (star), 20 (triangle), 50 (square) and 70 K (open circle)

The broad minimums in Figs. 7 (a,b) for nanoparticles with average sizes of 17 and 27 nm show larger contribution of the disordered surface shell to the anisotropy field in comparison with 94 nm nanoparticles. Consequently, we may suppose that the second minimum in Figs. 7 (a,b) at 70 K arises due to surface magnetic anisotropy. It confirms the existence of a core-shell magnetic structure in the $La_{0.7}Sr_{0.3}MnO_3$ nanopowders.

The singularities indicating the anisotropy field $H_{an}$ are clearly detectable. The observations displayed confirm that the applied method is an appropriate tool for measuring the magnetic anisotropy for this class of nanopowders. As the confirmation, the temperature dependences of magnetic anisotropy field for $La_{0.7}Sr_{0.3}MnO_3$ nanopowders are displayed in Fig. 8. It is seen that the anisotropy field decreases with increasing temperature and this dependence is very similar to that reported in [12] for $La_{0.7}Sr_{0.3}MnO_3$ single crystals.

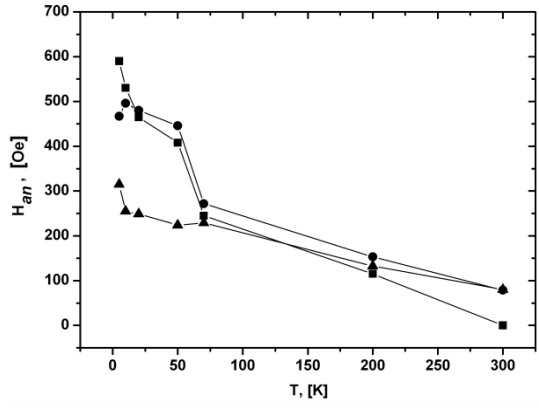

**Fig. 8**. The temperature dependence of the magnetic anisotropy field $H_{an}$ for the $La_{0.7}Sr_{0.3}MnO_3$ nanopowders annealed at 600 (square), 700 (circle) and 900 °C (triangle)

In order to determine the mechanism responsible for magnetic anisotropy in studied nanopowders the simplest way is to use the Callen and Callen model [13], which describes the temperature dependence of the anisotropy constant as a function of the saturation magnetization for single-ion mechanism:

$$\frac{K(T)}{K(0)} = \left(\frac{M_s(T)}{M_s(0)}\right)^{\frac{n(n+1)}{2}}, \qquad (3)$$

where $K(T)$ is the anisotropy constant at the temperature $T$, $K(0)$ is the anisotropy constant for $T = 0$ K, $n$ is the order of the spherical harmonics describing the angular dependence of the local anisotropy (n = 2 for uniaxial anisotropy, n = 4 for uniaxial and/or cubic anisotropies).

One should consider also contribution to the anisotropy constant due to dipolar (or pseudodipolar) two-ion interactions:

$$K_2 = cm^2 + fm^3 + km^{10}, \qquad (4)$$

$$H_{an} = am + bm^2 + lm^9, \qquad (5)$$

$a, b, c, f, k, l$ are the constants, $m = M_s(T)/M_s(5K)$.

The equation (5) describes two main contributions to the anisotropy field. The first term describes two-ion interactions, second and third terms are due to uniaxial anisotropy described in frames of the single-ion mechanism predicted by Callen and Callen model. The analysis of the experimental data were performed by fitting experimental results to the equation (5). The results of the fitting are presented in Fig. 9. These results confirm applicability of the used model of the magnetic anisotropy.

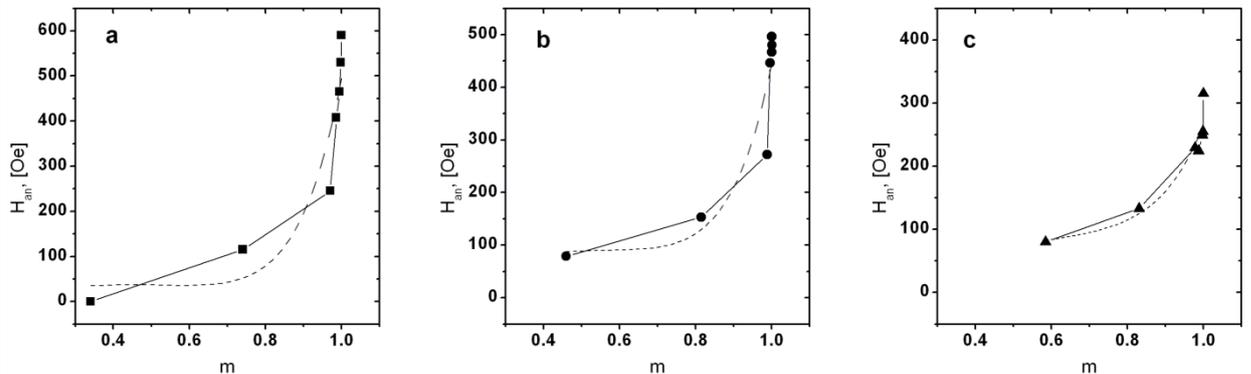

**Fig. 9.** The dependences of the magnetic anisotropy field on m = $M_s(T)/M_s(5K)$ for the $La_{0.7}Sr_{0.3}MnO_3$ nanopowders annealed at 600 (a), 700 (b) and 900 °C (c). Fitting according to Eq. 5 is marked by dashed lines

Table 1 presents the fitting parameters used in described procedure. Non monotonic dependence of the *a* and *b* parameters on particle size indicates on two different contributions to their magnitude.

**Table 1.** The values of fitting constants

| Annealing temperatures, [°C] | a, [Oe] | b, [Oe] | l, [Oe] |
|---|---|---|---|
| 600 | 171 | -200 | 524 |
| 700 | 334 | -317 | 427 |
| 900 | 207 | -115 | 176 |

Since Eq. (5) describes effective anisotropy it should consists of surface and bulk contributions with different dependence on *d* and *T*.

The anisotropy constant $K_2$ for different size of particles was calculated using relation (2).

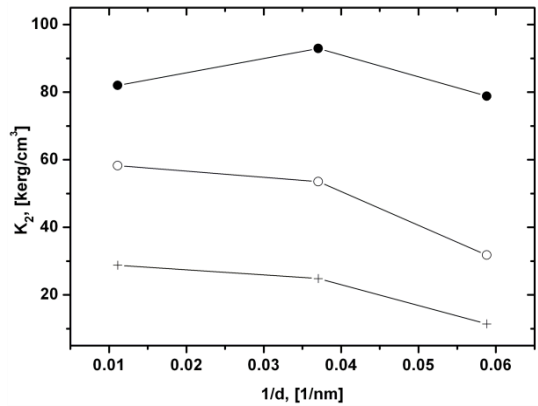

**Fig. 10**. The dependences of the anisotropy constant on the inverse particles size for the $La_{0.7}Sr_{0.3}MnO_3$ nanopowders, annealed at 600, 700 and 900 K, at experimental temperatures: 5 (circle), 70 (open circle) and 200 K (cross)

Usually, it is assumed that the effective anisotropy constant $K_2$ scales as the inverse of the particle's diameter *d* according to the relation [15]:

$$K_2 = K_{bulk} + 6K_s/d, \qquad (6)$$

where $K_{bulk}$ is the anisotropy constant for bulk manganite and $K_S$ is the surface anisotropy.

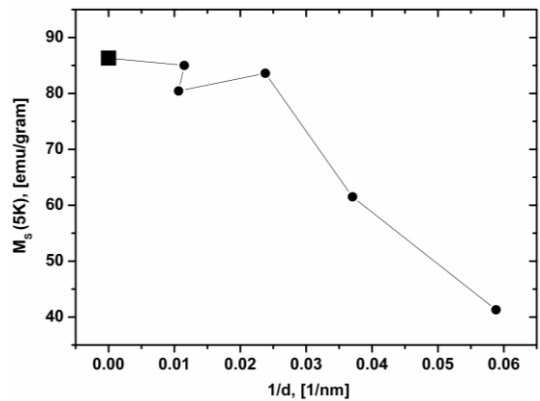

**Fig. 11.** The dependence of saturation magnetization on the inverse of particles size for the $La_{0.7}Sr_{0.3}MnO_3$ nanopowders annealed at 600, 700 and 900 K as well as at 800 and 1000 K taken from [5]. The corresponding dependence for bulk material ($d\to\infty$) at 5 K are marked by square [14]

The anisotropy constant and saturation magnetization are presented in Figs. 10 and 11 as a function of $1/d$. It is seen from Fig. 10 that the anisotropy constant is not linear function of $d^{-1}$ in contrast to predictions of (Eq. 6). Some deviation from Eq. (6) may result from the observation in [16] that the bulk and surface anisotropies are not additive values.

The saturation magnetization increases with increasing the particles size in nanopowders annealed at different temperatures (Fig. 11). In this case the extrapolation $d\to\infty$ leads to the value very near to that obtained for bulk $La_{0.7}Sr_{0.3}MnO_3$ manganite (86.3 emu/g [14])

## 5. Conclusion

The magnetic anisotropy of $La_{0.7}Sr_{0.3}MnO_3$ nanopowders was measured as a function of temperature by the modified singular point detection technique. The magnetic anisotropy consists of bulk and surface contributions. The TEM observations indicate that the $La_{0.7}Sr_{0.3}MnO_3$ nanoparticles are structurally homogeneous without any core-shell structure. Simultaneously, performed magnetic measurements (magnetization and magnetic anisotropy) indicate on the existence of magnetic core-shell structure. Magnetic field may suppress this structure.

## Acknowledgements

This work was financially supported by European Fund for Regional Development (Contract no. UDA-POIG.01.03.01-00-058/ 08/00).